\documentclass
[aps,prc,twocolumn,showpacs,
groupedaddress,superscriptaddress,
nolongbibliography
]{revtex4-2}
\usepackage{graphicx}
\usepackage{amsmath} 
\usepackage{amssymb}
\usepackage{amsfonts}
\usepackage{mathrsfs}
\usepackage{dcolumn}
\usepackage{bm}
\usepackage{rotating} 
\usepackage{MnSymbol}
\usepackage{color}
\usepackage[dvipsnames]{xcolor}
\usepackage[active]{srcltx}
\usepackage{multirow}
\usepackage{caption}
\usepackage{comment} 
\usepackage{physics}
\usepackage{ulem,soul} 
\usepackage{cancel} 
\newcommand{\bs}[1]{\boldsymbol{#1}}
\newcommand{\si}[1]{\,\rm{#1}}
\newcommand{\sn}[1]{\times10^{#1}}
\newcommand{\nn}[2]{^{#1}\rm{#2}}
\begin{document}
\title{
Alpha-decay from $^{44}$Ti: Microscopic alpha half-live calculation using normalized spectroscopic factor}

\author{A. C. Dassie}
\affiliation{Instituto de F\'isica Rosario (CONICET-UNR), Ocampo y Esmeralda, Rosario 2000. Argentina.}
\affiliation{Facultad de Ciencias Exactas, Ingenier\'ia y Agrimensura (UNR), Av. Pellegrini 250, Rosario 2000. Argentina.}
\author{R. M. Id Betan}
\affiliation{Instituto de F\'isica Rosario (CONICET-UNR), Ocampo y Esmeralda, Rosario 2000. Argentina.}
\affiliation{Facultad de Ciencias Exactas, Ingenier\'ia y Agrimensura (UNR), Av. Pellegrini 250, Rosario 2000. Argentina.}

\date{\today}

\begin{abstract} 
\begin{description} 
  \item[Background] The microscopic description of alpha decay from the nucleons' degree of freedom involves a two-step process. The first consists of the clusterization of neutron and proton pairs; the second involves the tunneling process.
  \item[Purpose] A robust protocol for calculating the normalized spectroscopic factor (amount of clustering), as defined by Fliessbach, and its error is established and used for calculating the alpha-width for the $0^+$ states of the nucleus $^{44}$Ti.
  \item[Method] The Gamow Shell Model is used to calculate the structure part of the alpha-decay (spectroscopic factor), while the Gamow wave function determines the reaction part (single particle width).
  \item[Results] The conventional and normalized spectroscopic factors are calculated for the ground and excited $0^+$ states of $^{44}$Ti and the alpha-width and half-life of the excited states. A near alpha-threshold state has an alpha half-life of 5 $\mu$sec.
  \item[Conclusions] The normalization does not appreciably modify the ground-state clusterization, while the excited states do. The non-resonant continuum significantly increases the clustering of some of the excited states, particularly the $T=2$ state. The normalized formation amplitude looks like a single-particle wave function.  
\end{description}
\end{abstract}


\maketitle

\section{Introduction}\label{sec.introduction}
Alpha-decay, a phenomenon as enduring as quantum mechanics, remains a captivating research area with ongoing theoretical advancements \cite{volyaPhys.Rev.C2015,yangPhys.Rev.C2021}. Traditionally, its decay width has been calculated via the Thomas-Lane equation \cite{thomasProg.Theor.Phys.1954,laneRev.Mod.Phys.1958} involving penetrability and reduced width or through an expression linking spectroscopic factors and single-particle reduced widths \cite{arimaPhys.Lett.B1972,fliessbachJ.Phys.GNucl.Part.Phys.1976}. Both formulations stem from time-independent reaction theory. An alternative time-dependent framework for alpha-decay was developed by Mang \cite{mangZ.Physik1957}, exhibiting formal equivalence to the time-independent approaches \cite{zehZ.FurPhys.At.Nucl.1963}. The numerical applications showed to be far smaller compared with experimental results  \cite{mangAnnu.Rev.Nucl.Sci.1964}, even using large configuration model spaces \cite{haradaProg.Theor.Phys.1961,janouchPhys.Rev.C1983,dodig-crnkovicPhys.Scr.1988}. A major breakthrough was made by Fliessbach reinterpreting the spectroscopic factor. He argued that, due to the antisymmetrization, the cluster-channel wave function was not normalized \cite{fliessbachJ.Phys.G:Nucl.Phys.1977}. Shell model calculations incorporating Fliessbach prescription got significantly closer to the experimental alpha-width of the archetypal nucleus $^{212}$Po but still failed to reproduce it \cite{tonozukaNucl.Phys.A1979,lenziPhys.Rev.C1993,delionPhys.Rev.C2000,idbetanPhys.Rev.C2012}. On the other hand, the hybrid cluster plus shell model approach, complemented with Fliessbach normalization, does reproduce the $^{212}$Po alpha-width \cite{vargaPhys.Rev.Lett.1992,lovasPhysicsReports1998}.

It seems that the Shell Model alone cannot gain enough correlations to achieve the necessary clusterization. The elements involved in the alpha-width calculation are the single-particle model space, the four-body basis, the two-body interactions, and the Fliessbach normalization. Reference \cite{idbetanPhys.Rev.C2012} is the first attempt to cover these elements by extending the single-particle proton and neutron bases to the resonant continuum and considering the normalization of the channel decay. Although the asymptotic of the alpha formation amplitude was properly described, the alpha width was still smaller than expected \cite{idbetanPhys.Rev.C2012}. Reference \cite{dassiePhys.Rev.C2023} resumes this project by completing the single-particle bases with the non-resonant continuum and the two-body interaction in all nucleon-nucleon channels. The present work covers the calculation of the Fliessbach normalization by providing a robust protocol for calculating the normalized spectroscopic factor and its numerical error. The nucleus $^{44}$Ti is used as a benchmark since it has many $0^+$ states; besides, it is of interest in nuclear astrophysics \cite{coopermanNuclearPhysicsA1977}, for example, for understanding its formation through the $\alpha$-Ca reaction in core-collapse supernova environments \cite{2012Robertson}.

The paper is organized as follows. Section \ref{sec.formalism} gives the absolute width in terms of the normalized spectroscopic factor. Section \ref{sec.method} briefly defines the parameters for the system and introduces the protocol for calculating the normalized spectroscopic factor and its error. Sec. \ref{sec.snormalized} calculates the normalized spectroscopic factor for the $0^+$ states of $^{44}$Ti of Ref. \cite{dassiePhys.Rev.C2023}. The microscopic alpha-decay width of the excited states are being calculated in Sec. \ref{sec.results.halflives}. The effects of truncating the many-body Hilbert space are studied in Sec. \ref{sec.ths}. Finally, Sec. \ref{sec.conclusions} summarizes the results.

\section{Formalism} \label{sec.formalism}
The appropriate microscopic calculation of the alpha-decay width $\Gamma_L$ in terms of the single particle width $\Gamma^\mathrm{sp}_L$ is given by the Arima expression \cite{arimaPhys.Lett.B1972} with the normalized spectroscopic factor as defined by Fliessbach \cite{fliessbachZ.FurPhys.At.Nucl.1975}
\begin{equation}
    \Gamma_L = \mathcal{S}_L \Gamma^\mathrm{sp}_L \, ,
    \label{eq:arimaexpression}
\end{equation}
with
\begin{equation}
    \mathcal{S}_L = \int G_L^2(R) R^{2} dR \, .
    \label{eq:norm_s}
\end{equation}

The modified formation amplitude $G_L$ is expressed in terms of the conventional one $g_L$ and the norm kernel $\mathcal{N}_L$ as follows \cite{fliessbachZ.FurPhys.At.Nucl.1975,fliessbachNucl.Phys.A1976,beckAnnalsofPhysics1987,vargaNucl.Phys.A1992},
\begin{equation*}
   G_L(R)= \int  \mathcal{N}_L^{-1/2}(R,R') \; g_L(R')R'^2  \;dR' \, ,
\end{equation*}
with \cite{haradaProg.Theor.Phys.1961}
\begin{align*}
 g_L(R)&= \int d\Omega_R \int d\xi_\alpha \int d\xi_D \; \nonumber \\
	 & \Psi_{J M}\;  
 	 \mathcal{A} \left[  \phi_\alpha(\xi_\alpha) \; \Psi^D_j(\xi_D)\; Y_L(\hat{R}) \right]^*_{J M},
\end{align*}
and 
\begin{align*}
   & \mathcal{N}_L(R,R') =  \\
   & \langle \mathcal{A} 
       \frac{\delta(R_\alpha - R)}{R^2} \phi_\alpha 
       \left[ Y_L \Psi_j^D \right]_{JM} 
       |
       \mathcal{A} 
       \frac{\delta(R_\alpha - R')}{R'^2} \phi_\alpha 
       \left[ Y_L \Psi_j^D \right]_{JM}  
       \rangle \, .
\end{align*}
The parent wave function (w.f.) $\Psi_{J M}$ is a double-closed shell w.f.  $\Psi^D_{j m}$ plus two protons and two neutrons, as defined in Ref. \cite{dassiePhys.Rev.C2023}. The alpha particle w.f. $\phi_\alpha$ is defined as in Ref. \cite{idbetanPhys.Rev.C2012}, and $Y_{L M_L}$ represents the angular part of the relative alpha-daughter w.f.

The norm kernel is expanded using an equidistant set of orthonormalized shifted Gaussian functions denoted as $\tilde{F}_L(R,R_k)$, where $R_k=k\, \Delta R$ \cite{fliessbachZ.FurPhys.At.Nucl.1976a}. This expansion extends to the radial coordinate $R_{\rm{max}}=M\, \Delta R$. 

Expressing the modified formation amplitude in terms of the norm kernel eigenvalues $n_\nu$ and eigenfunctions $u^L_\nu(R)$ yields \cite{idbetanPhys.Rev.C2012} 
\begin{equation}
   G_L(R) = \sum_{\nu=1}^{\nu_\mathrm{max}} \; 
      n_\nu^{-1/2} \; u^L_\nu(R) \; g^L_\nu \, ,   
    \label{eq:renorm_g}
\end{equation}
with
\begin{equation}
   g^L_\nu = \int  u^L_\nu(R) \; g_L(R) \;  R^2 dR,
\end{equation}
and the corresponding spectroscopic factor is given by
\begin{equation}
   \mathcal{S}_L = \sum_{\nu=1}^{\nu_\mathrm{max}} \, s_\nu^L 
      \label{eq:renorm_s} \, ,
\end{equation}
with $\nu_\mathrm{max}$ to be determined, and
\begin{equation}
   s_\nu^L= \int R^2 
	\frac{\qty[u^L_\nu(R) \; g^L_\nu]^2}{n_\nu}  dR 
 \label{eq:renorm_snu} \, .
\end{equation}

The formation amplitude is expressed in terms of single-particle configurations as
\begin{equation}\label{eq:formampl}
   g_L(R) = \sqrt{8} \sum_{J_n\,J_p} \sum_{a\le b} \sum_{c\le d}
            b^{J_n\,J_p}_{j_a\,j_b\,j_c\,j_d}  
            (-1)^{\theta} 
            \frac{\hat{j}_a \hat{j}_b \hat{j}_c \hat{j}_d}{2}
            I^{J_n\,J_p}_{j_a\,j_b\,j_c\,j_d}(R)
\end{equation}
where $\theta=J_n+J_p-j_a-j_c-l_b-l_d+1$. This equation relates the formation amplitude to the amplitudes of different configurations within the parent state. The pair indexes $a \le b$ labels the two-neutron basis, similarly, $c \le d$ for protons. The coefficients 
$b^{J_n\,J_p}_{j_a\,j_b\,j_c\,j_d}=Z^J_{np}\, X^{J_n}_{ab}\, X^{J_p}_{cd} $ enclosed the amplitudes of the correlated two-neutron and two-proton states $X^{J_n}_{ab}$ and $X^{J_p}_{cd}$, as well as the parent four-body wave-function amplitude \cite{dassiePhys.Rev.C2023} 
\begin{equation}
    \Psi_{JM} = \sum_{J_n\,J_p} Z^J_{np}
       \qty[\Psi_{J_n} \Psi_{J_p}]_{JM} \, .
\end{equation}

The following expression defines the radial matrix elements $I(R)$,
\begin{align}
    \begin{split}
  I^{J_n\,J_p}_{j_a\,j_b\,j_c\,j_d}(R) 
      =& W\qty(l_a j_a l_b j_b; \frac{1}{2} J_n) 
         W\qty(l_c j_c l_d j_d; \frac{1}{2} J_p) \\
       & \int d\Omega_R \int d\vb*{\rho}_1 d\vb*{\rho}_2 d\vb*{\rho}_3
           \, \phi_\alpha(\rho_1\rho_2\rho_3) \\
    & \left[
   \mathbb{S}\qty[\varphi^*_{a}(\bs{r}_1)\varphi^*_{b}(\bs{r}_2)]_{J_n} \mathbb{S}\qty[\varphi_{c}(\bs{r}_3)\varphi_{d}(\bs{r}_4)]_{J_p}\right]_{LM_L}
    \end{split}
\end{align}
with $\varphi_a(\bs{r})=\frac{u_{n_al_aj_a(r)}}{r}\, Y_{l_am_a}(\hat{r})$, $W$ the usual Racah's coefficients, and $\mathbb{S}$ the symmetrization and normalization operator. This equation calculates the overlap integral between the single-particle states in nucleonic coordinates $\bs{r}$ and the alpha wave function in relative coordinates $\bs{\rho}$ \cite{idbetanPhys.Rev.C2012}.

The expressions of this section reduce to that of Ref. \cite{idbetanPhys.Rev.C2012} for the monopole interaction, i.e., $J_n=J_p=0$ and $Z^J_{np}=1$.

\section{Model and Method} \label{sec.method}
\begin{figure}[h!t]
\centering
  \includegraphics[width=0.6\columnwidth]{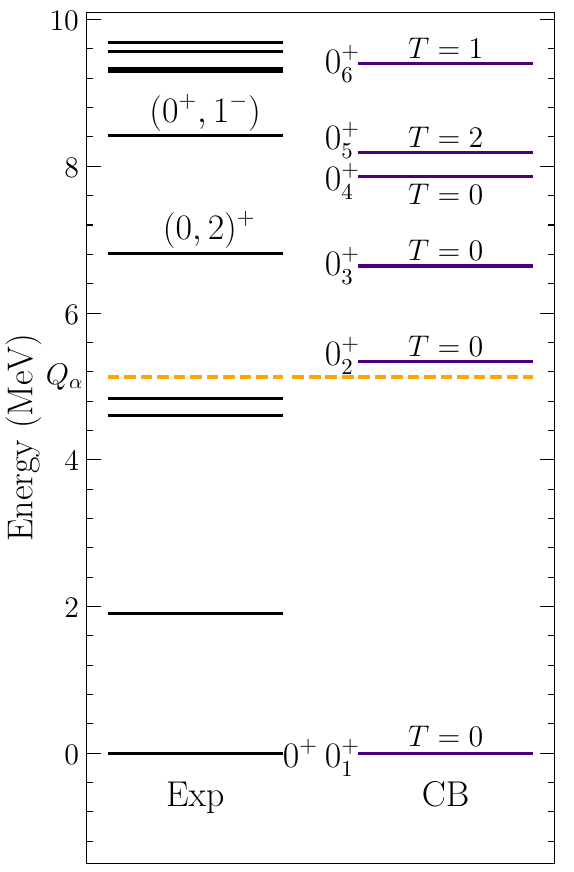}
  \caption{\label{fig:fbe_44Ti} Experimental (Exp) and calculated (CB) $0^+$ states of $\nn{44}{Ti}$ as explained in  Ref. \cite{dassiePhys.Rev.C2023}.}
\end{figure}

The six $0^+$ states of the $^{44}$Ti nucleus of the companion paper \cite{dassiePhys.Rev.C2023} are considered in the applications. These states are shown in Fig. \ref{fig:fbe_44Ti} for the complete basis. The pole (PB) and complete (CB) many-body basis are specified in Table III of Ref. \cite{dassiePhys.Rev.C2023}. The pole basis is a many-body representation that includes only the resonant part of the single-particle continuum. The complete basis also contains the complex energy scattering states associated with the resonances of the pole basis, plus the real energy scattering states for the other partial waves. The single-particle energies are calculated using the Woods-Saxon and Spin-Orbit mean fields in Table I of Ref. \cite{dassiePhys.Rev.C2023}. The interactions described in Section III.B of Ref. \cite{dassiePhys.Rev.C2023} determine the two-body energy levels. All these parameters determine the single particle wave function, the two- $X_{ab}^{J_n}$, $X_{cd}^{J_p}$, and four-body $Z^{0^+}_{np}$ amplitudes used for determining the formation amplitude. To study the impact of the four-body model space truncation on clusterization, we compare the calculation in the CB mentioned above with that of the truncated basis as defined in Ref. \cite{dassiePhys.Rev.C2023}.

For the first time, reference \cite{idbetanPhys.Rev.C2012} provided a thorough analysis of the impact of the parameters involved in determining the norm kernel and its effect on the normalized spectroscopic factors as defined by Fliessback \cite{fliessbachZ.FurPhys.At.Nucl.1975,fliessbachZ.FurPhys.At.Nucl.1976a}. The spacing $\Delta R$ between the Shifted Gaussian Functions (SGF) governs the quality of the basis set. If the SGF are widely separated, the basis becomes inadequate. If they are too close, it leads to an overcomplete representation \cite{saitoProg.Theor.Phys.1969}. Besides, the radial part of the SGF must also be normalized. The cutoff $R_{max}$ (or the number of normalized SGF, $M$) influences the final result. Finally, the cutoff on the norm kernel eigenvalue $\nu_{\rm{max}}$ significantly impacts the final value of the normalized spectroscopic factor. 

\begin{figure}[h!t]
\centering
 \includegraphics[width=0.8\columnwidth]{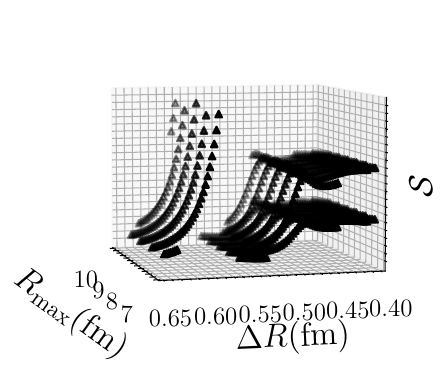}
  \caption{\label{fig:example} Schematic representation of the normalized spectroscopic factor of Eq. (\ref{eq:renorm_s}) as a function of $(\Delta R,R_{\rm{max}})$ parametrized on $\nu_{\rm{max}}$.}
\end{figure}

Let us qualitatively describe the protocol for obtaining a reliable normalized spectroscopic factor with an assigned error. The determination of the normalized spectroscopic factor is performed in two steps. First, we initialize the SGF parameters $\Delta R$ and $R_{\rm{max}}$ by expanding the single-particle core wave functions in the normalized SGF. A set of pairs $\Delta R$ and $R_\mathrm{max}$ is established by defining an error for the expansion of the wave functions and the normalization of the SGF. Then, the normalized spectroscopic factor $\mathcal{S}$, Eq. (\ref{eq:renorm_s}), is calculated for all possible $\nu_{\rm{max}} \leq M$. Many of the estimated $\mathcal{S}$ diverge, but some patterns with plateaus can also be found; see, for example, Fig. \ref{fig:example}. An exploration of this kind of figure gives a set of pair parameters $(\Delta R,R_{\rm{max}})$ of physical interest, as it is shown, for example, in Fig. \ref{fig:44Tiproject:3Dspectfactconvergence} in section \ref{sec.snormalized}.

Next, we must determine $\nu_{\rm{max}}$ to identify which plateau is adequate. The usual procedure is to take it as the change in the slope of $n_\nu$ versus $\nu$ \cite{fliessbachZ.FurPhys.At.Nucl.1976a,vargaNucl.Phys.A1992,lovasPhysicsReports1998,arimaAIPConf.Proc.1978,idbetanPhys.Rev.C2012}. This criterion is unstable for different values $(\Delta R,R_{\rm{max}})$, with $\Delta R$ and $R_{\rm{max}}$ in the plateau \cite{idbetanPhys.Rev.C2012}. Instead, we take all eigenvalues $n_\nu$ with the condition that $s_\nu^L < 1$. We found that this criterion is more stable and, in some cases, coincides with the change in the slope of $n_\nu$. The motivation for constraining $s_\nu^L < 1$ is that in practical applications, the most important contributions to $\mathcal{S}_L$ are given for the $s_\nu^L$ close to the slope change, just before $s_\nu^L$ becomes bigger than the unit.

To determine an error to the calculated $\mathcal{S}_L$, we thoroughly compute Eq. (\ref{eq:renorm_s}) on the three-dimensional parameter space $(\Delta R, R_{\rm{max}},\nu_{\rm{max}})$. This calculation provides a colossal amount of normalized spectroscopic factors organized in a histogram, see, for example Fig. \ref{fig:44Tiproject:histograms}.  Next, a Gaussian fit is performed around the calculated $\mathcal{S}_L$. The error is defined as the width $\sigma$ of the Gaussian distribution.

\section{Application: Amount of clustering} \label{sec.snormalized}
To illustrate the implementation of the protocol introduced in the previous section, the normalized spectroscopic factor is calculated for the $0^+$ states of $^{44}$Ti using the wave functions calculated in Ref. \cite{dassiePhys.Rev.C2023}. A critical discussion of the results, in comparison with experimental data, is also performed.

\begin{figure}[h!t]
	\centering
	\includegraphics[width=0.48\columnwidth]{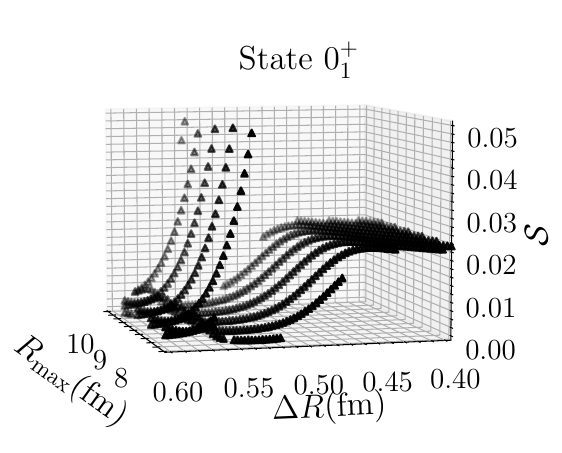}
	\includegraphics[width=0.48\columnwidth]{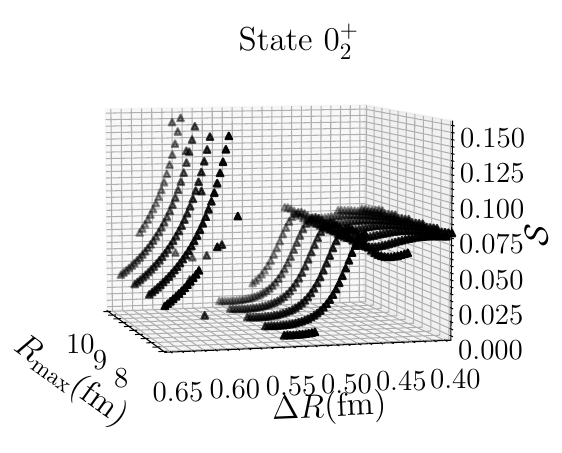}
	\includegraphics[width=0.48\columnwidth]{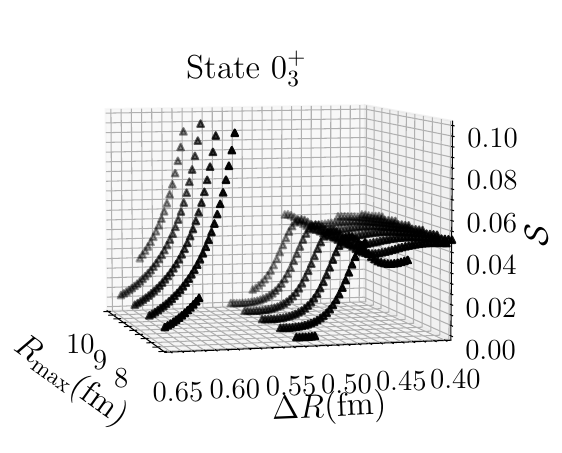}
	\includegraphics[width=0.48\columnwidth]{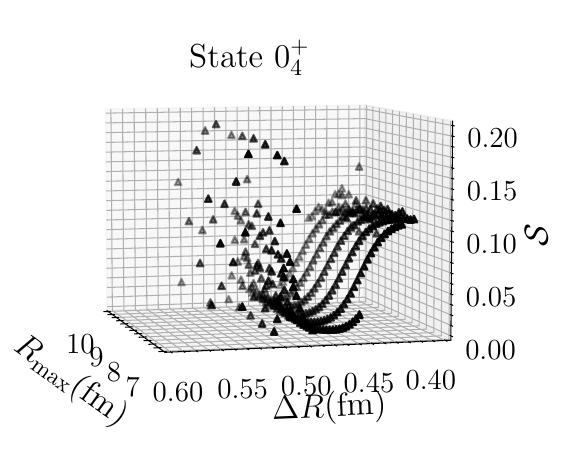}
	\includegraphics[width=0.48\columnwidth]{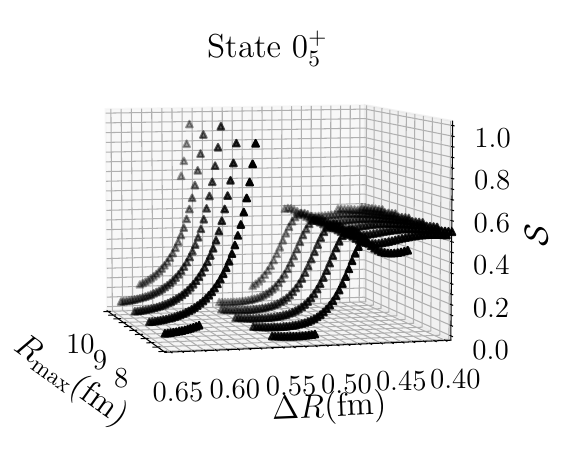}
	\includegraphics[width=0.48\columnwidth]{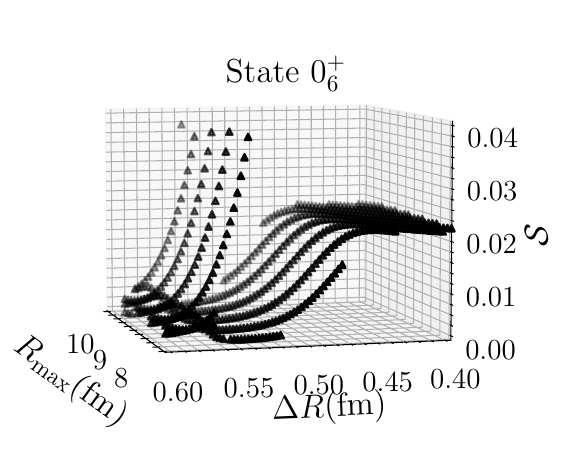}
	\caption{Normalized spectroscopic factor $\mathcal{S}$ parametrized in $\nu_{\rm{max}}$, as function of $\Delta R$ and $R_\mathrm{max}$.}
	\label{fig:44Tiproject:3Dspectfactconvergence}
\end{figure}

\textit{Determination of the parameters:} In order to estimate meaningful ranges for $\Delta R$ and $R_\mathrm{max}$, the single-particle core states of $^{40}$Ca are expanded in the normalized SGF. The mean-field for proton and neutrons that define these states are the ones in Table I of Ref. \cite{dassiePhys.Rev.C2023}. The error for the normalization of the SGF is $10^{-4}$, while the absolute limit for the expansion of the single-particle wave function is $10^{-2}$ fm$^{-1/2}$. With these constrains we get the following ranges $0.4\si{fm} \lesssim \Delta R \lesssim 0.8\si{fm}$ and $8\si{fm} \lesssim R_\mathrm{max} \lesssim 10\si{fm}$. Next, the normalized spectroscopic factor $\mathcal{S}$ is calculated for all possible values of $\nu_{\rm{max}}$, for each one of the $0^+$ states. The result of this calculation is shown in Figure \ref{fig:44Tiproject:3Dspectfactconvergence}.

\begin{figure}[h!t]
	\centering
	\includegraphics[width=0.48\columnwidth]{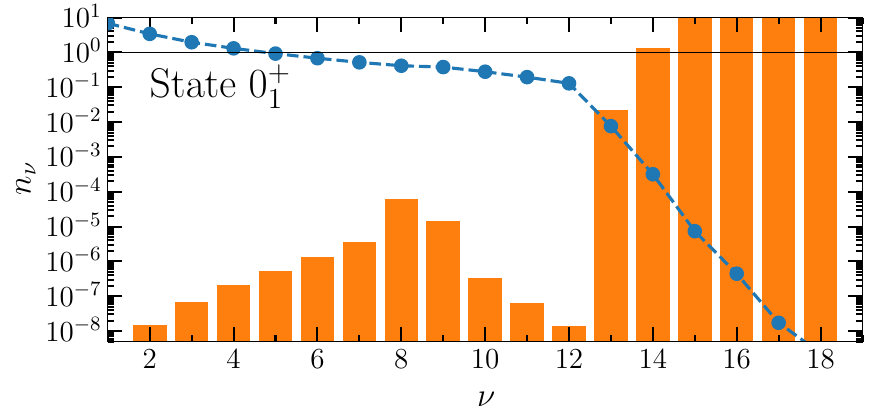}
	\includegraphics[width=0.48\columnwidth]{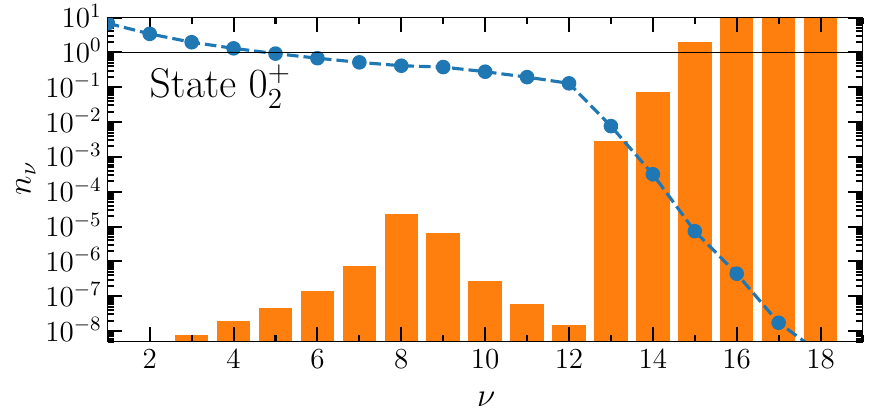}
	\includegraphics[width=0.48\columnwidth]{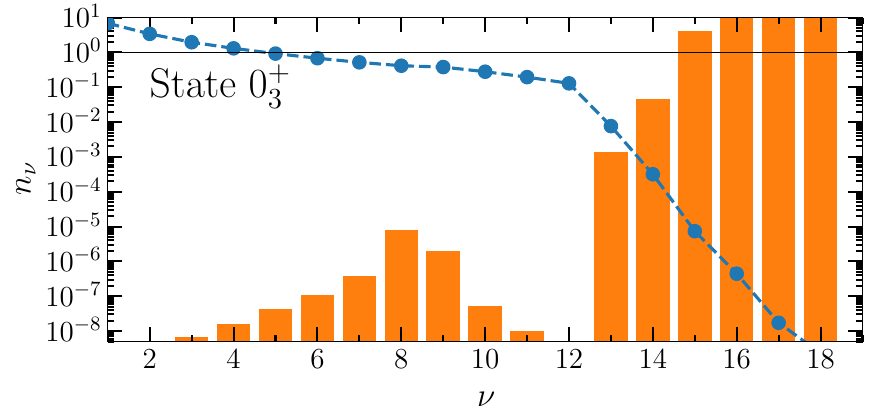}
	\includegraphics[width=0.48\columnwidth]{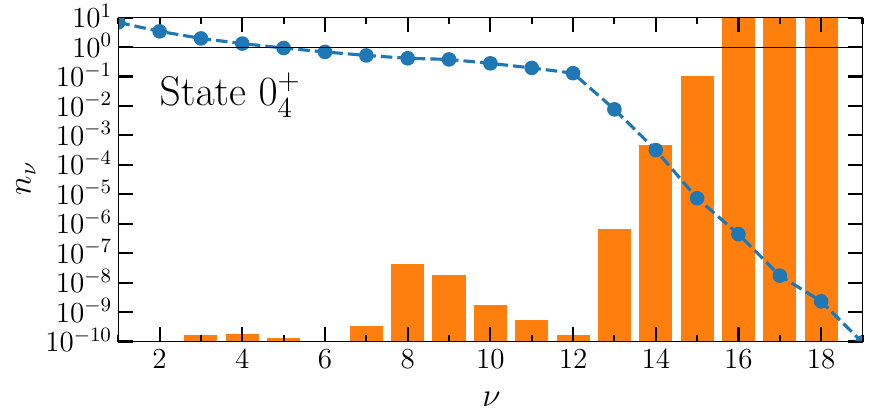}
	\includegraphics[width=0.48\columnwidth]{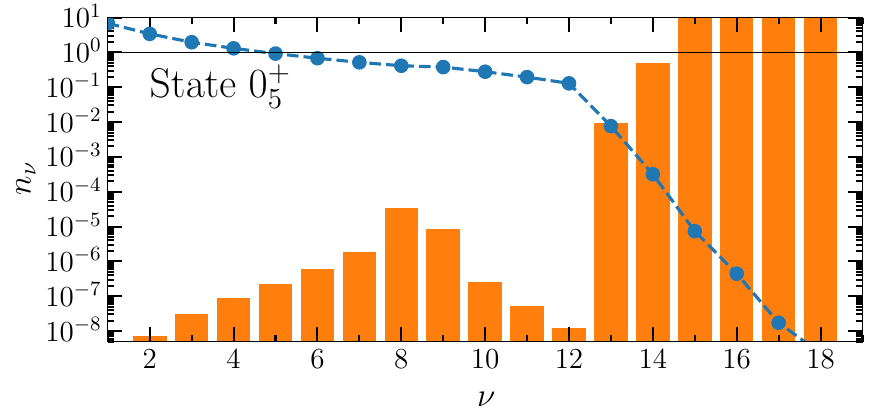}
	\includegraphics[width=0.48\columnwidth]{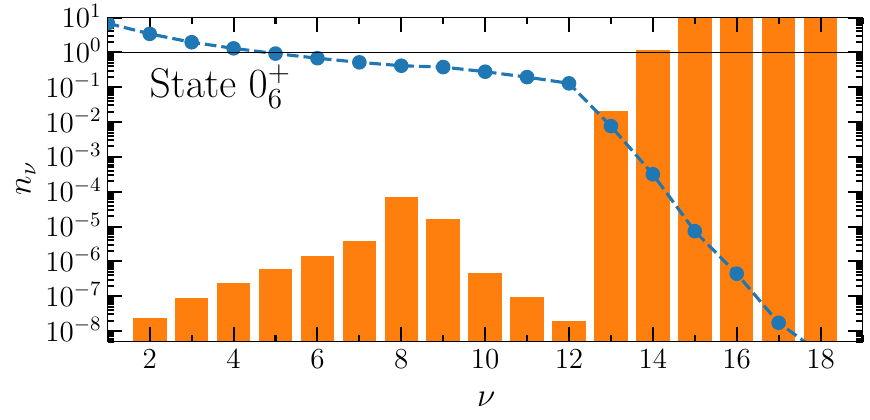}
\caption{\label{fig:44Tiproject:eigenvaluesversusnu} Norm kernel eigenvalues $n_\nu$ (filled circles) and partial spectroscopic factor $s_\nu$ (filled bars) using the complete basis for each one of the $0^+$ states of $^{44}$Ti.}
\end{figure}

The second step of the protocol requires determining the cutoff $\nu_{max}$ in Eq. \eqref{eq:renorm_s}. Figure \ref{fig:44Tiproject:eigenvaluesversusnu} shows the distribution of the norm kernel eigenvalues $n_\nu$ and $s_\nu$  for a typical case in the plateau region, $\Delta R=0.4$ fm, $R_\mathrm{max}=8$ fm (these values for $\Delta R$ and $R_\mathrm{max}$ are consistent with Refs. \cite{fliessbachZ.FurPhys.At.Nucl.1976a,fliessbachNucl.Phys.A1976}). The characteristic change in the slope of the norm kernel eigenvalues $n_\nu$ can be appreciated in the region for which $s_\nu$ became bigger than the unit $\nu \sim 15,\, 16$.  

\textit{Calculation of $\mathcal{S}$:} The calculated amount of clustering for each one of the $0^+$ states is shown in Table \ref{table:specfacts.fa_44Ti} for the pole (PB) and complete basis (CB). Due to the truncation of the non-resonant continuum in the PB, $\mathcal{S}$ has a spurious imaginary component which varies from 16 to 43\%. The inclusion of the non-resonant continuum (CB) wipes out the imaginary component and produces, in general, an enhancement of $\mathcal{S}$. The increase in the amount of clustering is remarkable for the $0^+_4$ and $0^+_5$ excited states. The influence of the non-resonant continuum in the amount of clustering contrasts with its effects on the wave function amplitudes (see Table I of the companion paper \cite{dassiePhys.Rev.C2023}), where the non-resonant continuum mildly changes its real part, but rectifies the spurious imaginary component. 
\begin{table}
\centering
\caption{\label{table:specfacts.fa_44Ti} Amount of clustering (error in parenthesis) for the ground and excited $0^+$ states of $^{44}$Ti calculated using the pole (PB) and complete (CB) bases.}
\begin{tabular}{c|c|c}
\hline
 & \multicolumn{2}{c}{$\mathcal{S} \times 10^{2}$} \\
 i & PB & CB \\
\hline
 $1$ & $1.7(2) -i\, 0.2$ & $2.1(2)$  \\
 $2$ & $5.1(2),-i\, 2.2$ & $7.3(4)$ \\
 $3$ & $6.4(3),-i\, 0.5$ & $4.6(2)$ \\
 $4$ & $1.2(6),-i\, 0.5$ & $10(1)$  \\
 $5$ & $25(1),-i\, 5$    & $49.3(2)$ \\
 $6$ & $1.8(5),-i\, 0.2$ & $2.1(1)$  \\
\hline
\end{tabular}
\end{table}

In Ref. \cite{dassiePhys.Rev.C2023}, it was assumed that clusterization could be inferred from the amount of collectivity of the wave function. So, only single-particle width was calculated for the states $0_3^+$ and $0_5^+$. The results of the amount of clustering of Table \ref{table:specfacts.fa_44Ti} reveal that this is not necessarily the case; for example, the state $0_4^+$ also has an appreciable clusterization.

\textit{States below the threshold:} No excited states below the alpha-threshold was found \cite{dassiePhys.Rev.C2023}. A shell model calculation using the effective Kuo and Brown interaction also found none of these experimental states \cite{1969Simpson,dixonPhys.Rev.C1978}. Presumably, because they have different configurations than four-nucleon valence states; in particular, the first excited state belongs to the $N=12$ band head \cite{michelPhys.Rev.Lett.1986}. The calculated spectroscopic factor of the ground state is 0.02. Since the spectroscopic factor is not an observable, it is model dependent \cite{2001Kramer,fulbrightNuclearPhysicsA1977,mazumderPramana-JPhys2016}. The ground state's alpha spectroscopic factor seems particularly intriguing since it ranges from 0.04 \cite{yamayaPhys.Rev.C1990}, 0.03-0.08 \cite{mazumderPramana-JPhys2016}, 0.2 \cite{yamayaPhys.Rev.C1993}, to $\sim$ 1 \cite{strohbuschPhys.Rev.C1974,kimPhys.Rev.C1992,guazzoniNuclearPhysicsA1993,fulbrightNuclearPhysicsA1977}. Our calculated value is closer to the one obtained from finite-range model \cite{mazumderPramana-JPhys2016} which uses complex squared Woods-Saxon optical potential \cite{1983Michel}.

\textit{States above the threshold:} The calculated amount of clustering of the state $0_2^+$ is only $0.07$. This state has no experimental counterpart but agrees in energy with Ref. \cite{shahPhys.Rev.1969}. In Ref. \cite{dassiePhys.Rev.C2023}, the calculated state at $6.641$ MeV was assigned to the $0_3^+$ due to its proximity to the experimental level $(0,2)^+$ at $6.810$ MeV \cite{bohneNucl.Phys.A1977}. The amount of clustering of this state is also small, 0.046. This figure compares well with 0.02 of Ref. \cite{yamayaPhys.Rev.C1990}, which assigns the quantum number $2^+$ to this state. The amount of clustering of the state $0_4^+$, at the energy 7.857 MeV, is 0.10. Even when this state is $\sim$ 700 keV apart, we associate it with the state $(0^+,1^-)$ (this state was inadvertently left out in the companion paper \cite{dassiePhys.Rev.C2023}) at the energy $8.54$ MeV. Reference \cite{strohbuschPhys.Rev.C1974} makes the tentative assignment of $0^+$ for the state $8.54$ MeV, with a tentative ratio of eight between its spectroscopic factor and the ground state. The amount of clustering of the state $0_5^+$ ($T=2$ \cite{dassiePhys.Rev.C2023}) is 0.49. Although 1.147 MeV apart from the experimental level at 9.338 MeV, it must be paired with it because it is the first experimental $T=2$ state. The last state, $0^+_6$, at the energy 9.401, has an amount of clustering of $0.02$. There are not experimental spectroscopic factors for $0^+$ states above 9 MeV, although, for high spin state they ranges from 0.02 to 0.003 \cite{yamayaPhys.Rev.C1990}.

\begin{figure}[h!t]
\centering
	\includegraphics[width=0.48\columnwidth]{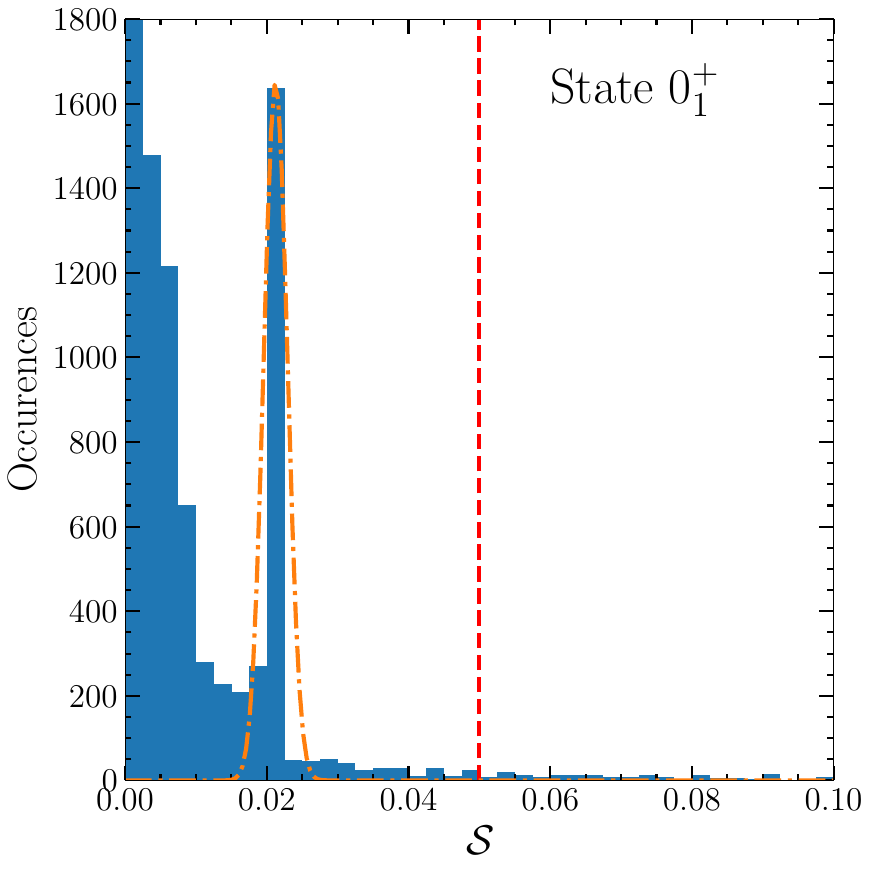}
	\includegraphics[width=0.48\columnwidth]{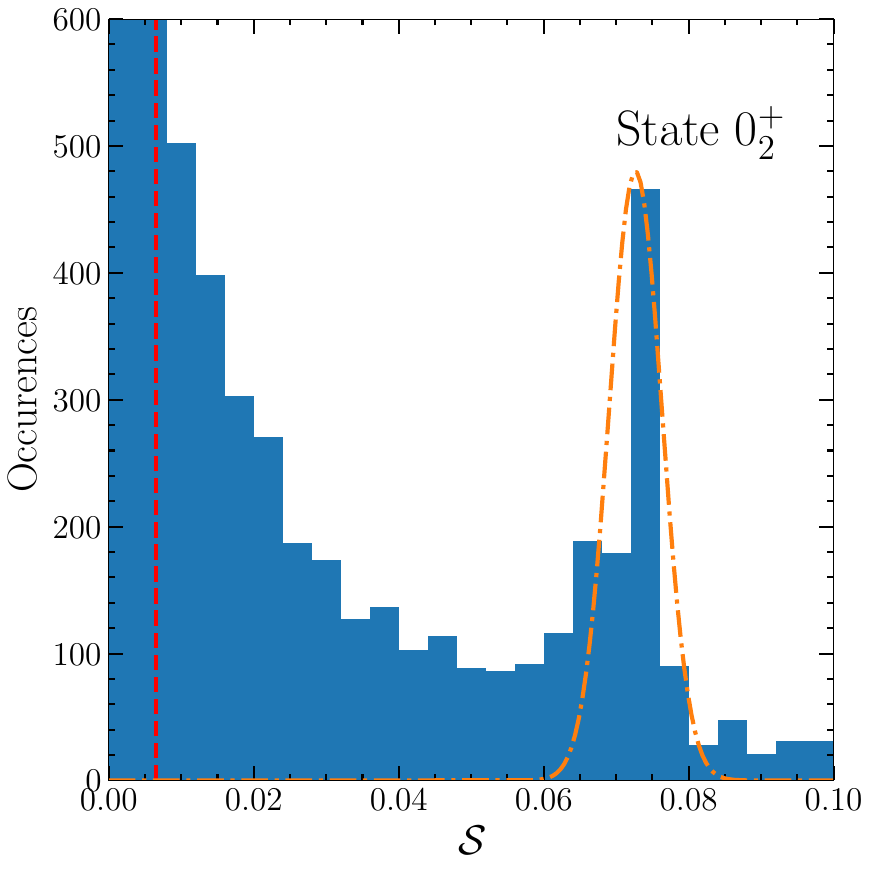}
	\includegraphics[width=0.48\columnwidth]{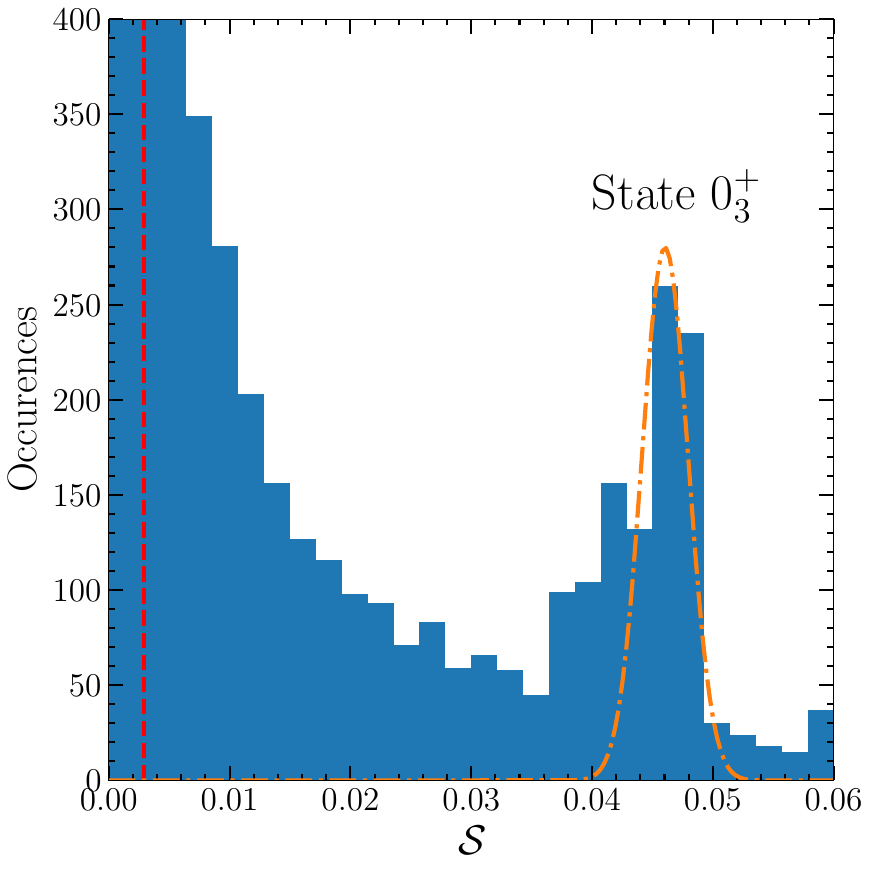}
	\includegraphics[width=0.48\columnwidth]{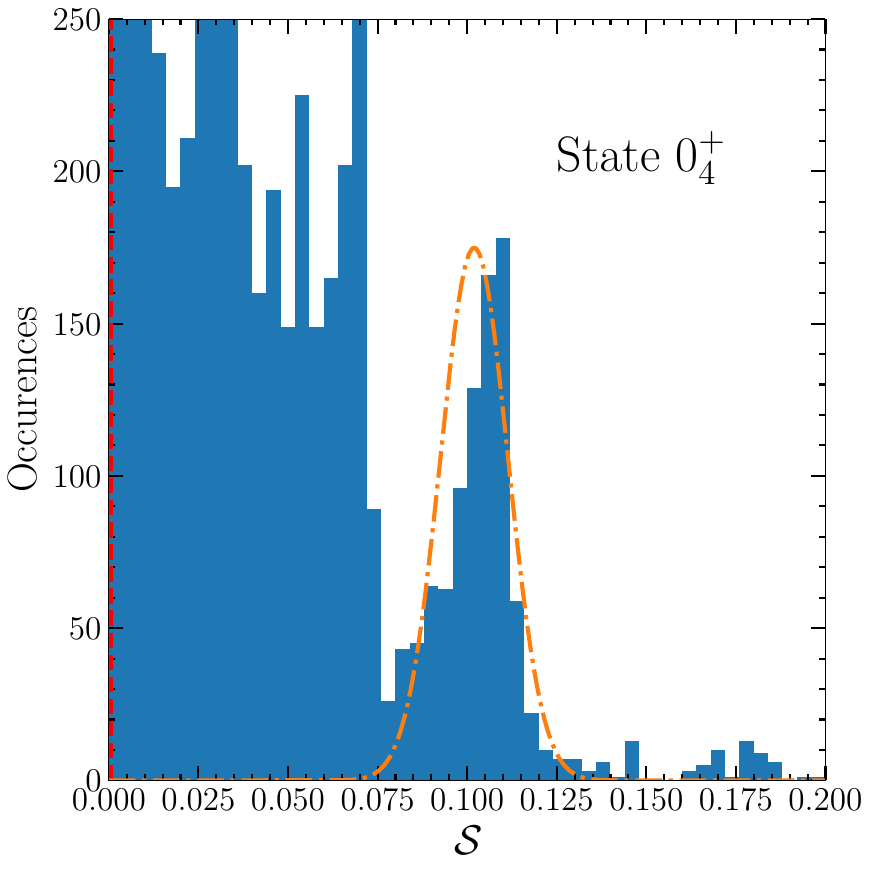}
	\includegraphics[width=0.48\columnwidth]{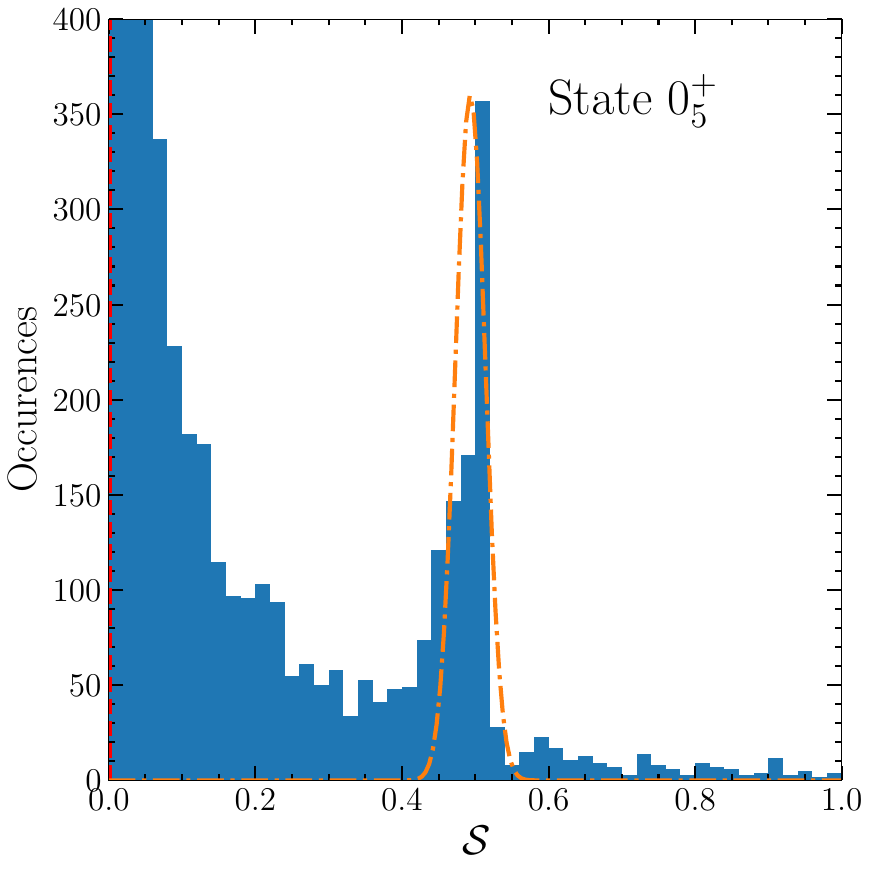}
	\includegraphics[width=0.48\columnwidth]{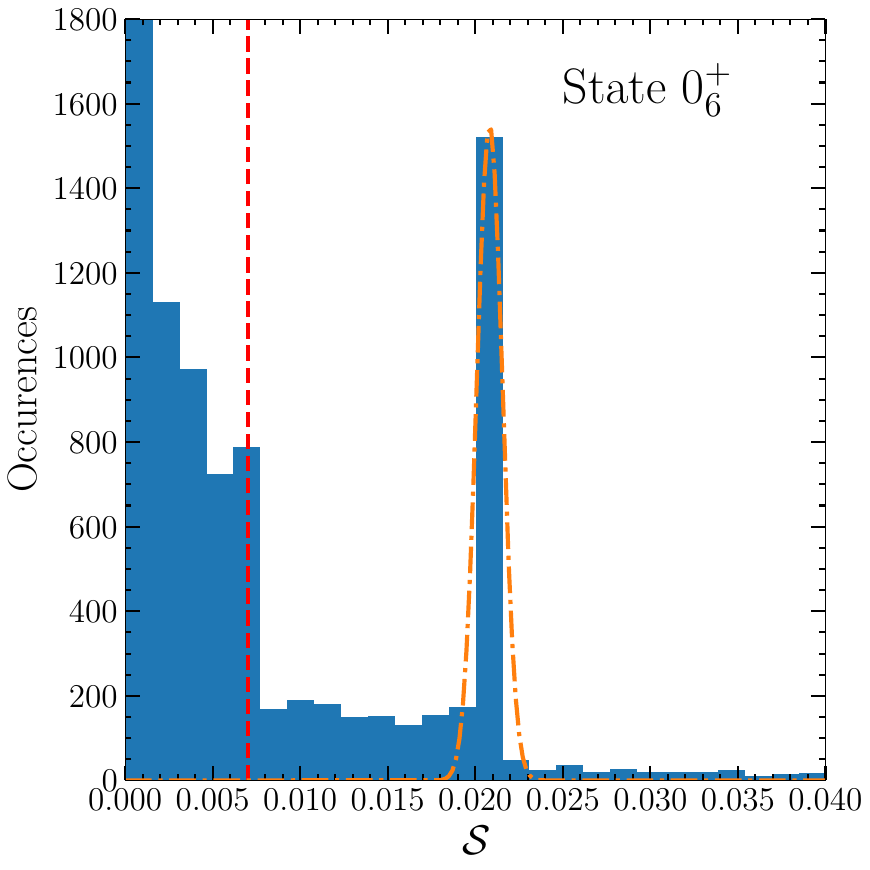}
\caption{\label{fig:44Tiproject:histograms} Histogram for occurrences of the normalized spectroscopic factors calculated as explained in Sec. \ref{sec.method}. The dashed vertical lines in each histogram correspond to the usual spectroscopic value \cite{idbetanPhys.Rev.C2012}. The fitted Gaussian curve is shown around each normalized spectroscopic factor of Table \ref{table:specfacts.fa_44Ti} (CB).}
\end{figure}

\textit{Error for $\mathcal{S}$:} Following the protocol described in Section \ref{sec.method} for calculating the error, we get the histograms shown in Fig. \ref{fig:44Tiproject:histograms}. The errors range between 5\% and 10\%,  as shown in Table \ref{table:specfacts.fa_44Ti}. The state $0^+_6$ exhibits the narrower Gaussian distribution when it corresponds to the state with a more extensive convergence plateau in Fig. \ref{fig:44Tiproject:3Dspectfactconvergence}. A dashed vertical line indicates the usual spectroscopic factor in the figure. For all states but the ground state, the Fliessbach procedure increases the spectroscopic factor.  

\begin{figure}[h!t]
\centering
\includegraphics[width=\columnwidth]{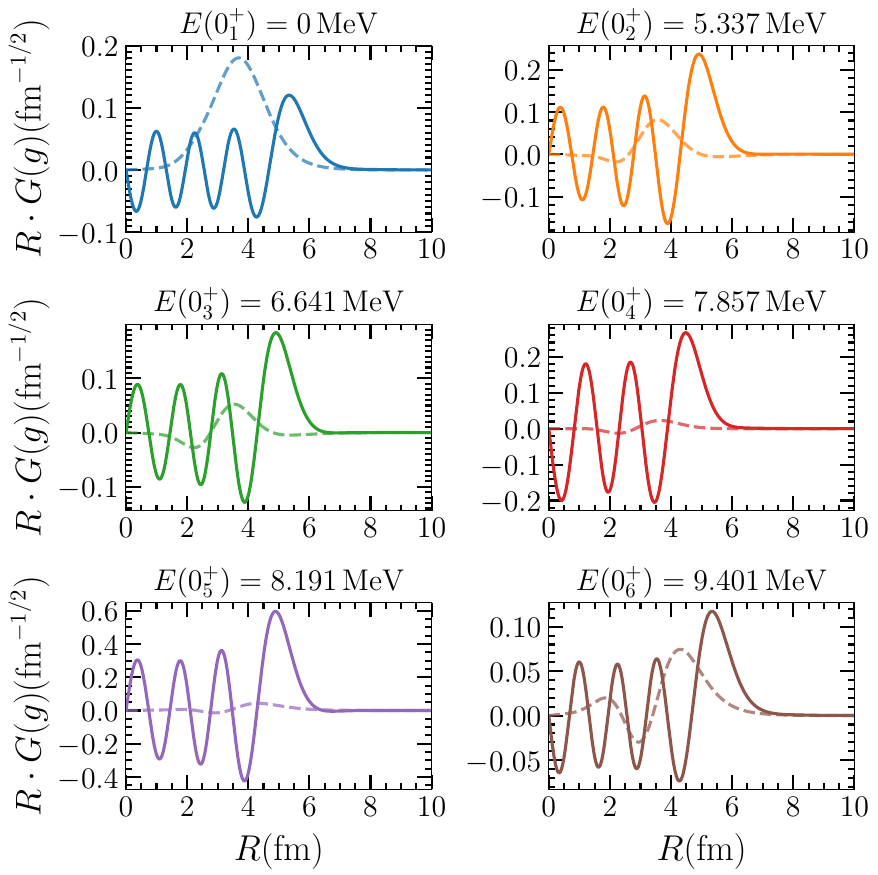}
\caption{\label{fig:fa} Conventional $g(R)$ (dashed line) and normalized $G(R)$ (continuum line) formation amplitudes of the six states $0^+$ calculated in the complete basis.}
\end{figure}

\textit{Conventional spectroscopic factor:} The usual $g(R)$ and normalized $G(R)$ formation amplitudes for the complete basis are shown in Fig. \ref{fig:fa}. The formation amplitude shows the usual overlap shape \cite{idbetanPhys.Rev.C2012}. In contrast, the normalized one resembles an alpha single-particle wave function with several nodes ranging from six to eight. The maximum of the normalized forming amplitude is noticeably larger than the maximum of the usual formation amplitude, except for the ground state. The Fliessbach normalization produces a displacement of the peak to the outer region of the nuclear surface around $\sim 5-6\si{fm}$, consistent with a rough approximation $R_{core}+R_\alpha \sim 6$ fm.

\section{Application: alpha-decay width and half-life} \label{sec.results.halflives} 
We calculate the single-particle width $\Gamma_{\rm{sp}}$ of the excited states by taking the average of the calculation obtained with the two mean-field potentials as explained in the companion paper \cite{dassiePhys.Rev.C2023}. Using the amount of clustering $\mathcal{S}$ of the previous section (Table \ref{table:specfacts.fa_44Ti}), we get the absolute alpha-width $\Gamma=\mathcal{S} \, \Gamma_{\rm{sp}}$ and half-life $T_{1/2}=\ln 2 \, \hbar / \Gamma$. The error for the single-particle width was defined through the two mean fields as discussed in Ref. \cite{dassiePhys.Rev.C2023}, while the error for $\Gamma$ corresponds to the error propagation from $\Gamma_{\rm{sp}}$ and $\mathcal{S}$. These results are presented in Table \ref{table.NewGammas}.
\begin{table}[h!t]
\centering
\caption{\label{table.NewGammas} Alpha single-particle width $\Gamma_{\rm{sp}}$ and alpha width $\Gamma$ scaled with the amount of clustering $\mathcal{S}$, for the excited states of $^{44}$Ti in the complete basis \cite{dassiePhys.Rev.C2023}. The calculated half-live of the $0_5^+$ state incorporated $\Gamma_\gamma=0.75\times 10^{-6}$ MeV \cite{freedmanPhys.Rev.C1978}. Errors are given in parentheses.}
\begin{tabular}{c|ccc}
\hline
  State 
  & $\Gamma_\mathrm{sp}$ (MeV) & $\Gamma$ (MeV) & $T_\mathrm{1/2}$ (sec) \\
\hline
  $0_2^+$ 
  & $0.13(2)\sn{-14}$ & $0.95(15)\sn{-16}$ & $0.48\sn{-5}$  \\
  $0_3^+$ 
  & $0.42(16)\sn{-12}$ & $0.19(7)\sn{-13}$ & $0.24\sn{-7}$ \\
  $0_4^+$ 
  & $0.11(1)\sn{-5}$ & $0.11(1)\sn{-6}$ & $0.43\sn{-14}$ \\
  $0_5^+$ 
  & $0.15(4)\sn{-4}$ & $0.74(20)\sn{-5}$ & $0.56\sn{-16}$ \\
  $0_6^+$ 
  & $0.54(10)\sn{-2}$ & $0.11(2)\sn{-3}$ & $0.40\sn{-17}$ \\
\hline
\end{tabular}
\end{table}

\begin{figure}[h!t]
\centering
\includegraphics[width=\columnwidth]{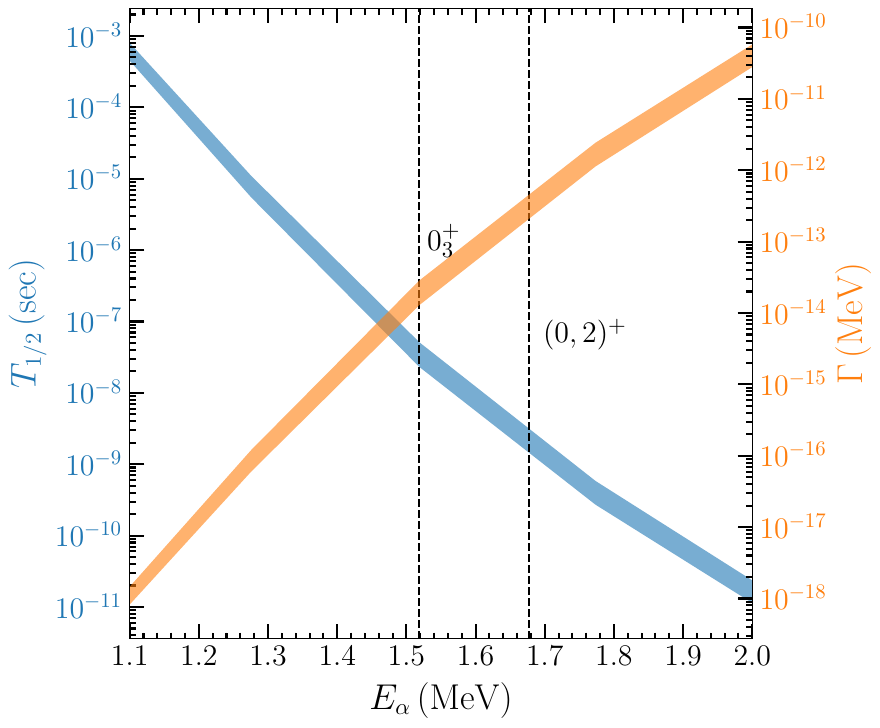}
\caption{\label{fig:gammavsE} Alpha-decay width (increasing curve with right scale) and half-life (left scale) of the state $0^+_3$ as functions of the decay energy. The calculated energy and the energy of the paired estate are shown as vertical lines. The width of the curves represents the error calculated, as explained in the text.}
\end{figure}

The calculated half-life of the not observed $0_2^+$ state is of the order of microsecond due to its proximity to the alpha-threshold, while the half-life of the $0_3^+$ is $24$ ns, three times bigger than the lower limit proposed in Ref. \cite{dassiePhys.Rev.C2023}. Figure \ref{fig:gammavsE} shows that its half-live decreases by a factor of ten when the energy is taken as the paired state $(0,2)^+$. Vertical lines in the figure indicate the calculated and experimental states.

It is known that the $T=2$ state decay mainly by $\gamma$ but also by $\alpha$, with $\Gamma_\gamma=0.75\pm 0.19$ eV and $\Gamma_\alpha=0.35\pm 0.07$ eV \cite{freedmanPhys.Rev.C1978}. Our calculated value is $7.4$ eV, which is twenty times bigger than the experimental one. This could indicate that the clusterization is not as big (49 \%) as our result predicts. The $0^+_4$ state can be considered an alpha-decay candidate for $^{44}$Ti with a half-life of $4.3$ fs.

\section{Truncated Hilbert Space}\label{sec.ths}
The Gamow wave function and the Gamow Shell Model for calculating $\Gamma_{\rm{sp}}$ and $\mathcal{S}$, respectively, unify the reaction and structure parts of the alpha-decay calculation. There are two main parts of this unified approach which demand considerable computation resources. The first is calculating the four-body wave function in the full Berggren representation, including the non-resonant continuum. The second one is the many-dimensional integrals. To reduce the computation time, we investigate the effect of the many-body Hilbert space truncation on the amount of clustering. We calculate it in the four-body basis generated with the neutron-neutron and proton-proton bases up to 10 MeV of excitation energy, called \textit{Case II} in Ref. \cite{dassiePhys.Rev.C2023}. Table \ref{table.SoverS} shows the result.
\begin{table}[h!t]
\centering 
\caption{\label{table.SoverS} Ratio of the amount of clustering in the CB versus the one in the truncated basis (\textit{Case I} and \textit{Case II}, respectively in Ref. \cite{dassiePhys.Rev.C2023}).} 
\begin{tabular}{c|cccccc}
\hline
  Ratio & $0_1^+$ & $0_2^+$ & $0_3^+$ & $0_4^+$ & $0_5^+$ & $0_6^+$ \\
\hline
  $\frac{\mathcal{S_{\textit{Case I}}}}{\mathcal{S}_{\textit{Case II}}}$
  & $2.6$ 
  &  $50$ 
  &  $15$ 
  & $12$ 
  & $1.3$ 
  & $1$ \\ 
\hline
\end{tabular}
\end{table}

The amount of clustering decreases in most of the states calculated on the truncated basis. This indicates that high-energy excited states of the many-body bases are important to build the clusterization. This behavior contrasts with the one found for the wave function amplitudes, for which the truncation has no significant effects (see Table VIII in \cite{dassiePhys.Rev.C2023}.) This analysis shows that truncating the many-body Hilbert space may be justified for calculating spectra and wave functions but not for calculating the amount of clustering.

\section{Summary and conclusions} \label{sec.conclusions}
The present work and the companion paper \cite{dassiePhys.Rev.C2023} takes the $0^+$ states of the nucleus $^{44}$Ti to refine the technique introduced in Ref. \cite{idbetanPhys.Rev.C2012} for the calculation of the half-live of drip-line nuclei. The resonant model space of Ref. \cite{idbetanPhys.Rev.C2012} was complemented with the non-resonant continuum. Consequently, the spurious imaginary part on the two- and four-body wave functions and on the formation amplitudes disappeared. An effective Gaussian interaction replaced the schematic separable force. This force produced more realistic two-particle spectra and wave functions. The missing proton-neutron interaction in \cite{idbetanPhys.Rev.C2012} was incorporated into the formalism. This extra interaction permitted adjustments to the four-body threshold. It also contributes to gaining extra correlations, which are essential in the alpha-decay process. Finally, a protocol for calculating the normalized spectroscopic factor and an associated numerical error was introduced. Besides all these improvements, the calculation indicates that the tensor interaction is essential for properly describing the neutron-proton threshold and possibly affects the amount of clustering.

It was shown that the many-body Hilbert space truncation, although valid for determining the spectrum and wave functions, is not valid for clustering calculation. This result indicates that while the effect of the missing correlations in the spectrum may be compensated by renormalizing the interaction, this is not the case when it comes to clustering many nucleons. It was also shown that the non-resonant continuum is important not only to eliminate the spurious imaginary part of the pole approximation but also to create essential multi-nucleon correlations. 

The Fliessbach normalization produces that the profile of the spectroscopic factor changes from an overlapping shape to that of the alpha-daughter wave function. The normalized spectroscopic factor can then be used to generate an effective Heel-Wheeler potential between the alpha cluster and the core nucleus. The effect of Fliessbach normalization is also noticeable in a shift of the maximum of amplitude towards the outside part of the radial coordinate.

The amount of clusterization of the calculated $0^+$ states of $^{44}$Ti range from $\sim 2\%$ to $\sim 50\%$. The value of $\mathcal{S}$ of the ground state is small, in agreement with some bibliography and strong disagreement with others. This particular state shows a significant model dependence on the definition of the spectroscopic factor. For the excites states, the minimum ($\sim 2\%$) is reached by the $T=1$ state, while the maximum is for the $T=2$ one, although this figure may be overestimated since the calculated width is twenty times bigger than the experimental one. We found a near-threshold alpha-decay state with a half-live of $5\, \mu\rm{s}$. The calculated half-live for the state $0^+_3$ paired with the experimental $(0,2)^+$ is the order of nanoseconds, while the state $0_4$, paired with the experimental $(0^+ 1^-)$, is calculated to have a half-live of $4$ fs.

As a final remark, including the non-resonant continuum allows alpha-decay calculation beyond the drip line. In particular, in the region $N=Z=50$, where each proton's single-particle state is unbound. An application to the alpha-decay of the nucleus $^{104}$Te is in progress.

\begin{acknowledgments}
This work has been partially supported by CONICET (Consejo Nacional de Investigaciones Científicas y Técnicas, Argentina) through Grant No. PIP-0930. The computations were performed on the Computational Center of CCT-Rosario and CCAD-UNC, members of the SNCAD, MincyT-Argentina.
\end{acknowledgments}

%

\end{document}